\documentclass[11pt]{spieman}  
\usepackage{amsmath,amsfonts,amssymb}
\usepackage{graphicx}
\usepackage{setspace}
\usepackage{tocloft}
\usepackage{lineno}
\usepackage{rotating}
\usepackage{soul}

\title{Characterization of the C-RED 2: A High Frame Rate Near-Infrared Camera}

\author[a,b,*]{Rose K. Gibson}
\author[b]{Rebecca Oppenheimer}
\author[c]{Christopher T. Matthews}
\author[c]{Gautam Vasisht}
\affil[a]{Columbia University, Department of Astronomy, 550 West 120th Street, New York, USA, 10027}
\affil[b]{American Museum of Natural History, Department of Astrophysics, Central Park West at 79th Street, New York, USA, 10024-5192}
\affil[c]{Jet Propulsion Laboratory, California Institute of Technology, 4800 Oak Grove Dr, Pasadena, USA, 91109}

\cftpagenumbersoff{figure}
\cftpagenumbersoff{table} 
\graphicspath{{./}{./figures/}}
\begin{document} 
\maketitle
\begin{abstract}
A new wave of precision radial velocity instruments will open the door to exploring the populations of companions of low mass stars. 
The Palomar Radial Velocity Instrument (PARVI) will be optimized to detect radial velocity signals of cool K and M stars with an instrument precision floor of 30 cm/s.  
PARVI will operate in the $\lambda = 1.2-1.8$ $\rm{\mu m}$ wavelength range with a spectral resolution of $\lambda/\Delta\lambda$ $\sim$100,000.
It will operate on the Palomar 5.1 m Hale telescope and use Palomar's PALM-3000 adaptive optics system, single-mode fibers, and an H band laser frequency comb to probe and characterize the population of planets around cool, red stars. 
In this work we describe the performance of the PARVI guide camera: a C-RED 2 from First Light Advanced Imagery. 
The C-RED 2 will be used in a tip-tilt loop which requires fast readout at low noise levels to eliminate any residual guide errors and ensure the target starlight stays centered on the fiber. 
At -40$^{\circ}$ C and a frame rate of 400 FPS in non-destructive read mode, the C-RED 2 has a combined dark and background current of 493 $e^-$/s. 
Using up-the-ramp sampling we are able to reduce the read noise to 21.2 e$^-$.  
With the C-RED 2, PARVI will be able to guide using targets as faint as 14.6 H magnitude.
\end{abstract}


{\noindent \footnotesize\textbf{*}Rose K. Gibson,  \linkable{rose.gibson@astro.columbia.edu} }


\section{Introduction}
\label{sec:intro} 
The search for exoplanets has captivated astronomers since their discovery in the early 1990s \cite{1995Natur.378..355M,1992Natur.355..145W}. With the advent of {\it Kepler} the number of confirmed exoplanets increased by orders of magnitude and transformed exoplanet studies. While much has been done in exoplanet science, the focus for the most part has been on sun-like stars. Of the 4,057 confirmed exoplanets (as of October 6th, 2019), 60.1\% of those are around G stars or earlier, and only 5.3\% orbit M stars \cite{2014PASP..126..827H}. Cool stars are the most common stars in the galaxy\cite{2003PASP..115..763C,1955ApJ...121..161S} and are expected to host one or more rocky planet each\cite{2015ApJ...807...45D}, and yet we know the least about their planetary systems. 

One reason for this extreme discrepancy is that the most productive exoplanet missions, such as {\it Kepler} and the High Accuracy Radial velocity Planet Searcher (HARPS) \cite{2004SPIE.5492..148R}, are best suited to observe sun-like stars. {\it Kepler} and its successor {\it K2} were sensitive to the 400 - 900 nm wavelength range \cite{2010ApJ...713L..79K}, and HARPS has a similar range of 383 - 690 nm. M stars range from temperatures of 3823 to 2088 K \cite{1996ApJ...461L..51B} which correspond to peak wavelength emissions between 758 and 1388 nm. These cool stars are much dimmer in optical bandwidths compared to sun-like stars, and therefore require significantly more observing time to achieve a similar signal-to-noise. To efficiently observe K and M dwarfs one must go to the NIR where the bulk of their spectral information is\cite{2010ApJ...710..432R}. An advantage to searching for low mass exoplanet hosts is that it is easier to detect a radial velocity (RV) signal from low mass stars due to low mass planets. This is simply due to the much higher mass ratio of the star-planet system. The high mass ratio means that the star is further from the system's barycenter and thus has a larger gravitational reflex motion due to the presence of the companion. In addition, sun-like stars are found to have less RV jitter in the NIR\cite{2015ApJ...798...63M}, a pattern which may extend to later type stars, with the caveat of star spot dependence\cite{2009A&A...498..853R, 2010ApJ...710..432R}.

The Palomar Radial Velocity Instrument (PARVI) is part of a next generation of instruments that will shift exoplanet hunting into the near-infrared. There are many new precision RV instruments coming online soon; a comprehensive list of new spectrographs can be found in \citenum{2017RNAAS...1a..51W} and those expected to operate beyond 1 $\rm{\mu m}$ are highlighted in Table \ref{tab:apra}. PARVI is expected to have notably superior instrumental velocity precision due to using single-mode fibers (SMFs) and a laser frequency comb (LFC) for wavelength calibration.

Recently, astronomers have invested in developing methods to use SMFs in combination with extreme adaptive optics (AO) to enhance ground based telescope performance\cite{2014Sci...346..809C,2014SPIE.9147E..7PJ,2016PASP..128l1001J}. 
Extreme AO systems, such as PALM-3000\cite{2013ApJ...776..130D} (P3K), provide diffraction-limited light that can be coupled to small SMFs with Strehl ratios of $\sim 80\%$ in the H band.  
SMFs provide a number of advantages over multi-mode fibers, such as eliminating focal ratio degradation and modal noise, and ensuring a spatially and temporally stable point spread function (PSF)\cite{2016PASP..128l1001J}. In addition to providing stable PSFs, SMFs divorce the resolution of a spectrograph from the physical parameters of the telescope. Instead, the resolving power depends solely on the slit size (fiber size) and chosen optics of the spectrograph design. This allows for much smaller spectrographs which in turn are easier to mechanically and thermally stabilize.

PARVI, to be installed on Palomar's 5.1 m Hale telescope, will produce  high-resolution spectra ($\lambda/\Delta\lambda$ $\sim$100,000) over the 1.2-1.8 $\rm{\mu m}$ wavelength range.  The primary science goal of this instrument is to conduct a 3-year survey of nearby K and M dwarf stars to search for exoplanets and other companions. With an instrument velocity precision floor of 30 cm/s, we will be able to detect objects less massive than the Earth around K and M dwarfs.  
In addition to the survey, PARVI will be used for Transiting Exoplanet Survey Satellite (TESS) candidate follow-up observations and to monitor variability of low-mass stars and brown dwarfs.  PARVI is also an exercise in testing new technology and observing methods, such as using extreme AO in combination with SMFs, a fiber-fed spectrograph, and a novel LFC for wavelength calibrations.

A critical component to achieving the precision necessary for these observations is the fine guiding system. The guiding system on the telescope, as described in detail in Section \ref{sec:instrument}, requires a fast readout detector to maintain the alignment of the beam on the single-mode fibers, which then send the light to the spectrograph.  For the guiding system, we will use a First Light Advanced Imagery C-RED 2 camera which contains a SNAKE-SW Indium Gallium Arsenide (InGaAs) detector.  In this work we describe a characterization of the C-RED 2 and discuss its expected performance as the PARVI guide camera. In Section \ref{sec:instrument} we outline the design of PARVI. In Section \ref{sec:methods} we describe the tests we performed with the C-RED 2. In Sections \ref{sec:snr} and \ref{sec:discus} we discuss the outcomes of those tests and the expected performance of the C-RED 2 on sky given our characterization.

\section{Instrument Design}
\label{sec:instrument} 
A unique aspect of PARVI is the use of single-mode fibers. 
Two other spectrographs are planning on using SMFs in combination with AO: iLocator \cite{2016SPIE.9908E..19C} and IRD \cite{2018SPIE10702E..11K}.
Traditionally, multi-mode fibers are used for fiber-fed spectrographs, such as SPIROU \cite{2012SPIE.8446E..2RM} and HARPS \cite{2004SPIE.5492..148R}, because they are larger ($\sim$100 micron core diameter) and therefore accommodate seeing-limited PSFs. 
The plate scale of PARVI's 9 $\rm{\mu m}$ diameter SMF is 10.2 mas/$\rm{\mu m}$, or 91.8 mas across the fiber. 
At PARVI's longest wavelength the diffraction limited beam is 90.7 mas across and fully contained within the SMF core.  
Because PARVI is fiber fed with SMFs, we eliminate flexure due to gravity vector changes and atmospheric effects, and significantly reduce mechanical and thermal noise. 

Another frontier in precision spectroscopy is the use of laser frequency combs (LFCs) for wavelength calibration. LFCs create precise, evenly-spaced spectral lines that are stable at the kilohertz level over the course of years, which translates to $<1$ cm/s drift over PARVI's lifetime \cite{2015arXiv150706344B}. 
PARVI will use an H band LFC with equal frequency spacings at 10 GHz, which creates thousands of lines across the full bandpass. 
The calibration lines and science data will be imaged on the spectrograph detector simultaneously. 

\subsection{Optical Path}
The light from the adaptive optics system P3K enters the fiber injection unit bench where $5\%$ of it is diverted to the C-RED 2 for guiding. 
The bulk of the light is sent to a multi-fiber ferrule, where one fiber sends the science light to the spectrograph, a second sends sky-background, and a third is back injected to create a false star image on the C-RED 2.
The guide science light and false star are simultaneously imaged on the C-RED 2 so a fast steering mirror can correct out AO residuals, non-common optical tilt, and flexure. 
The plate scale on the C-RED 2 detector is 1.02 mas/$\mu$m, and the pixel size is 15 $\mu$m, so the maximum point spread function is contained in a 5$\times$5 pixel square. 
A proportional-integral-derivative loop uses the centroid position of the science light on the detector relative to the reference stimulus to adjust the tip/tilt mirrors on the optical bench. 
The loop aligns the science light on the SMFs within 1 $\rm{\mu m}$ of the 9 $\rm{\mu m}$ fiber core. 
The science and sky-background light are directed through $\sim$100 m of fibers to the spectrograph off-telescope.

\subsection{C-RED 2}
The C-RED 2 is a low noise, fast readout camera with a SOFRADIR SNAKE-SW focal plane array. A full presentation of the camera design expected performance can be found in \citenum{2018SPIE10703E..1VF} and is the main comparison for our characterization tests. The array is made of 640$\times$512 pixels of size 15$\times$15 $\mu$m and is sensitive to wavelengths 0.9 - 1.7 $\rm{\mu m}$ at 70\% quantum efficiency. The camera does not require cryogenic cooling. Instead, a thermoelectric cooler is used to maintain temperatures as low as -40$^{\circ}$ C on the detector. There is an internal fan capable of dispersing the built-up heat, however, as recommended by First Light, we used a Koolance ERM-3K3UC chiller to remove excess heat from the C-RED 2 and avoid using the fan, which could introduce vibration to the camera and optical bench. The full array can be read at 400 frames per second\footnote{Firmware versions 2.9.1 and later allow for 600 FPS full frame readouts.} (FPS), or subarrays can be read on the order of thousands of reads per second. 

Images are recorded using a correlated double sampling (CDS) mode or an integrated multiple readout (IMRO) mode. At 400 FPS the gain in CDS and IMRO modes is 2.28 and 1.70 $e^-$/analog digital unit (ADU), respectively, with a 14-bit analogue-to-digital converter. In CDS mode the pixels are sampled after the detector reset and at the end of a single full read. The sampling is done directly on the C-RED 2 sensor. The nominal exposure time in this mode ranges from 50 $\mu$s to 1/FPS.  IMRO mode is a non-destructive read (NDR) mode where detector can be read $N$ times during the course of one exposure. One exposure yields $N$ reads, each recording the charge accumulation at the time of the $N^{\text{th}}$ read, with the exception of the first read which is a CDS image. The detector is read at the frequency of the frames per second for $N = 2$ to $N = 256$ reads. The exposure time of the initial CDS frame can be set to anything between 50 $\mu$s to 1/FPS. 
We characterized the behavior of the camera using both CDS and IMRO configurations in order to determine which configuration will be best for all observing situations.

We obtained a C-RED 2 for PARVI on April 2$^{\text{nd}}$, 2018 and conducted testing to evaluate the claims from the manufacturer as well as to to determine how it will perform as a fast guiding camera. The C-RED 2 was installed in the Astrophysics lab at the American Museum of Natural History. We used a Matrox Radient eV-CL frame grabber and a standard Camera Link connection with both Ubuntu (lab testing) and CentOS 7 (production setup). We used the Matrox Imaging Libraries (MIL) software and {\tt Python} wrapper included with the hardware to command the frame grabber. First Light provided front-end demo software that used MIL to control the C-RED 2. We used both the First Light demo software and our own {\tt Python} code to grab and save images from the camera for this characterization.

\section{Characterization Tests}
\label{sec:methods} 
\subsection{Bias Level}
The bias level on the detector is set by a bias voltage applied to each pixel, the semiconductor properties, and the specifics of the read-out electronics. Because there is pixel-to-pixel variation, the bias level across the detector is non-uniform.  This results in near-Gaussian, fixed pattern noise in every image. Here we calculate the bias level of the C-RED 2 and compare it to a theoretical Gaussian noise distribution. 

\begin{figure}[h]
\begin{center}
\begin{tabular}{c}
\includegraphics[width=.7\linewidth]{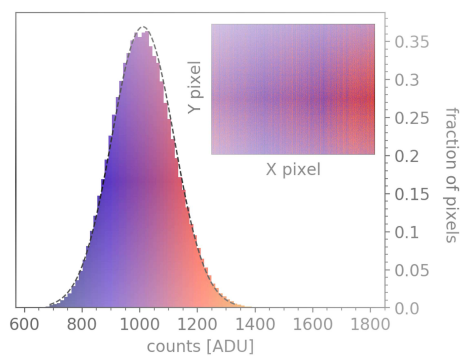}
\end{tabular}
\end{center}
\caption 
{\label{fig:bias} An average of 10 full array bias images (insert) taken at 400 FPS, $-40^{\circ}$ C and a histogram of the fraction of pixels at any given count value.
A Guassian centered on 1011 ADU and standard deviation 108 ADU is plotted on top of the histogram (black dashed line). 
The colors in the image correspond to the count number in the histogram.
The counts range from 571 to 1850 ADU.
} 
\end{figure}

\begin{figure}
\begin{center}
\begin{tabular}{c}
\includegraphics[width=\columnwidth]{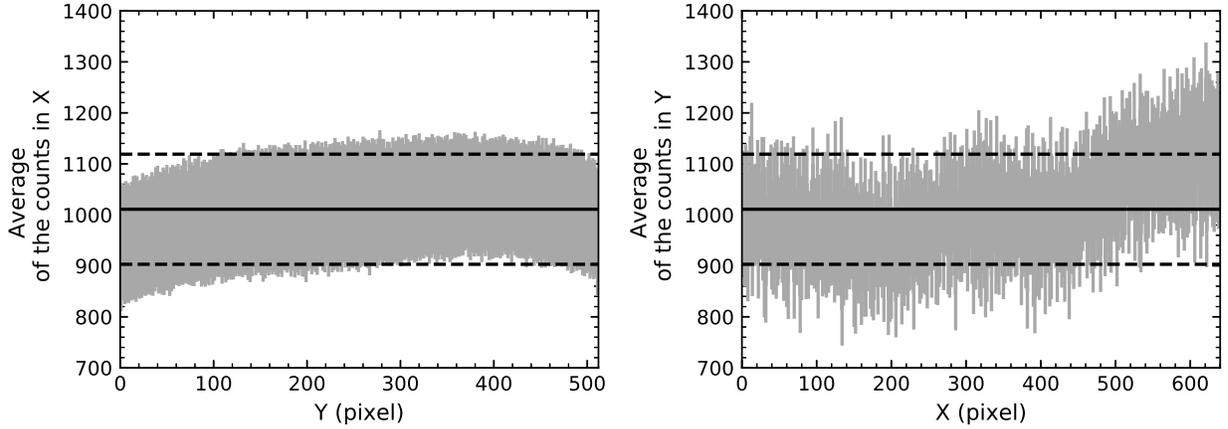}
\end{tabular}
\end{center}
\caption 
{ \label{fig:biasspat}
Average and standard deviation of each pixel in rows Y (top) and columns X (bottom) in a bias image. 
Plotted in black is the average counts (solid) plus and minus one standard deviation (dashed) for the entire detector ($\mu = 1011$ ADU and $\sigma = 108$ ADU).
} 
\end{figure}

We took a series of single, full array, dark CDS images at a constant bias voltage\footnote{Using First Light's latest firmware (2.9.1 and later) users can adjust the bias voltage. During our characterization we used an older firmware version and did not have that ability, therefore for every test mentioned here we used the default voltage (2V).}, 400 FPS, and $T=-40^{\circ}$ C. 
The average bias from ten images is 1011 $\pm$ 108 ADU (2305 $\pm$ 248 e$^-$ using gain of 2.28 $e^-$/ADU). 
The average image and corresponding histogram are show in figure \ref{fig:bias}, including a Gaussian profile centered on 1011 ADU with a standard deviation of 108 ADU.
The residuals of the true bias values and the Gaussian distribution deviate by $0.006\%$.
There is structure in the spatial distribution, as evident by visually inspecting the bias frame, and upon further inspection of the bias levels across the rows and columns of the detector (Figure \ref{fig:biasspat}). The average counts per row are effectively within one sigma the full detector average and standard deviation, however the average counts per column are more widely distributed and better contained within two sigma.

\subsection{Read Noise}
\label{subsubsec:readnoise}
\begin{figure}[h]
\begin{center}
\begin{tabular}{c}
\includegraphics[width=.8\columnwidth]{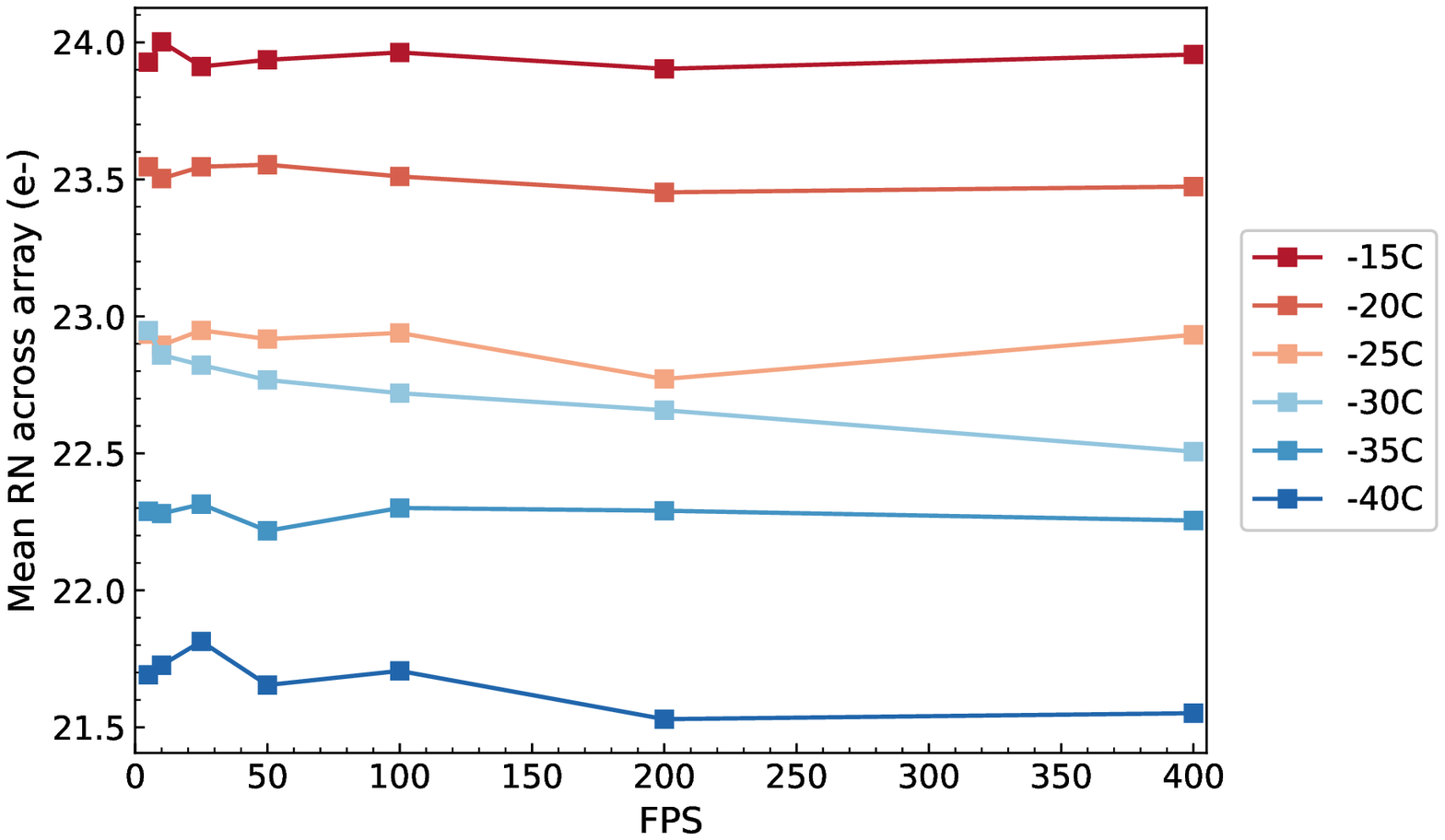}
\end{tabular}
\end{center}
\caption 
{ \label{fig:rnVfps}
Average single read CDS read noise [$e^-$] across the C-RED 2 detector at different framerates and temperatures. Each data point is the average noise from 10 bias images at $t_{\text{int}}=20$ $\mu$s. The lowest noise is 21.6 e$^-$ at FPS = 400 and $T = -40^{\circ}$C.
} 
\end{figure}

The read noise of the instrument is determined by a number of factors both on and off the detector. It can be quantified as the amount of variation the detector has between identical read outs. To measure the read noise of the C-RED 2 we took 10 single, full frame short exposures and calculated the standard deviation of each pixel over the 10 identical exposures. We then averaged the $640\times 512$ variations to determine the average read noise across the detector. We did this for dark images with frame rates of 400, 200, 100, 50, 25, 10, and 5 FPS and temperatures of $-40, -35, -30, -25, -20,$ and $-15^{\circ}$ C with an exposure time of 20 $\mu$s. The results are plotted in Figure \ref{fig:rnVfps}. The read noise increases with temperature, which is consistent with Figure 5 in \cite{2018SPIE10703E..1VF}. Using a FPS of 400 at a temperature of $-40^{\circ}$ C yields the smallest read noise of 21.6 $e^-$/pixel.

\subsection{Linearity}
\label{sec:linearity}

\begin{figure}
\begin{center}
\begin{tabular}{c}
\includegraphics[width=\linewidth]{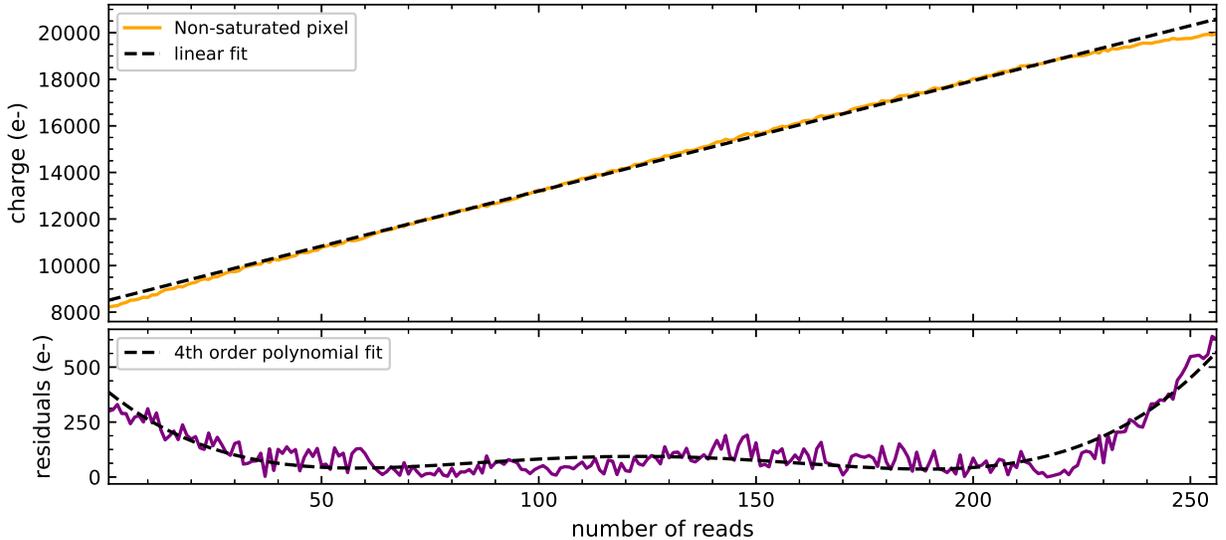}
\end{tabular}
\end{center}
\caption 
{ \label{fig:linear}
{\it Top}: The detector response (solid orange) and linear fit to the charge accumulation (dashed black) versus number of reads.
{\it Bottom}: Residuals between charge accumulation and linear fit (solid purple) and a 4th order polynomial fit to the residuals (black dashed). The RMS of the residuals to the polynomial fit is $0.738$ e$^-$.
The configurations for this exposure were FPS = 400, T = $-40^{\circ}$C, full frame, evenly illuminated detector. 
} 
\end{figure}

We tested the detector linearity with an IMRO exposure and fit a line to the charge accumulation over the exposure.  We used a framerate of 400 FPS, T = $-40^{\circ}$ C, evenly illuminated detector, and full image read out. We fit a line to a non-saturated pixel's charge accumulation, shown in Figure \ref{fig:linear}. The bottom panel of the figure shows the residuals between the measured charge and the first order linear fit.  The average residual is $0.42\%$ with a maximum of $3.98\%$ near the beginning and end of the exposure. This is within the 4$\%$ range published by SOFRADIR.

\begin{figure}[h]
\begin{center}
\begin{tabular}{c}
\includegraphics[width=\linewidth]{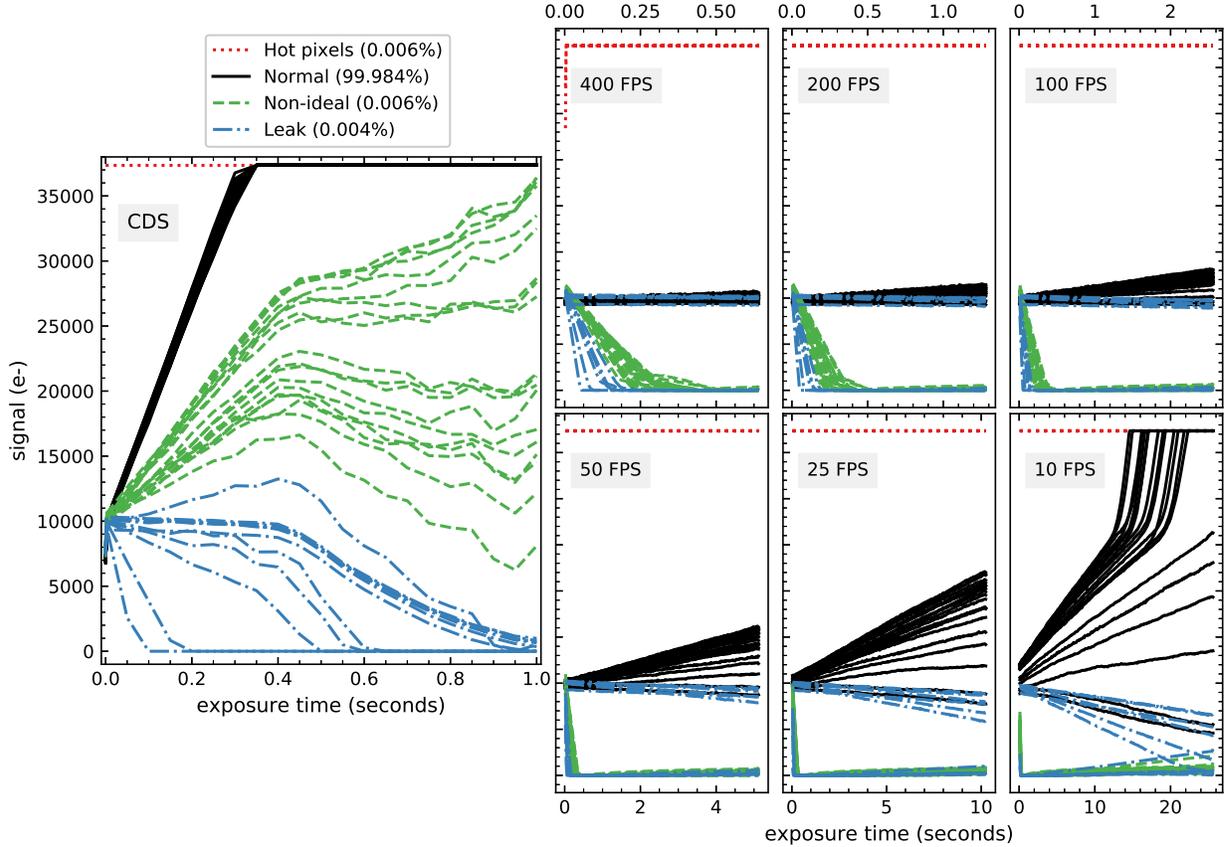}
\end{tabular}
\end{center}
\caption 
{ \label{fig:pixelcounts}
{\it Left:} Charge accumulated on a subset of pixels from 22 full frame CDS exposures at T = -40$^{\circ}$ C. We placed the pixels in one of four categories: hot (dotted red), normal (solid black), non-ideal (dashed green), or leak (dashdot blue). Because there are so few, all of the hot, non-ideal, and leak pixels are plotted. For the normal pixels, only the first 100 on the detector are shown.
{\it Right:} Charge accumulation across the C-RED 2 detector in IMRO mode for the same set of pixels at FPS = $400, 200, 100, 50, 25$ and 10.  Each panel  shows the charge counts from a single exposure with 256 non-destructive reads. 
} 
\end{figure}

We also examined the pixel behavior beyond the well-behaved linear pixels.  We took similar data sets in IMRO and CDS mode to compare the behavior and noise of the detector in each mode. For the CDS images we used an integrating sphere and piece of Teflon over the camera opening to evenly illuminate the pixels on the detector.   The detector temperature for each exposure was $-40^{\circ}$ C. We took one full frame image  with the following exposure times: $t_{\text{int}}$ = 0.005, 0.252, 5, 10, 15, 20, 25, 30, 35, 40, 45, 50, 55, 60, 65, 70, 75, 80, 85, 90, 95, and 100 ms. Figure \ref{fig:pixelcounts} follows a sample of pixels over the 22 exposure times.  The pixel grouping represents the different behaviors we found on this detector: hot pixels, normal, non-ideal, and leak. The hot pixels are those that saturate either immediately or very quickly.  These are the same in both CDS and IMRO mode. The normal pixels are those that accumulate charge until they saturate and remain at that level. The non-ideal and leaking pixels are those that may accumulate charge but also seem to lose charge at longer exposures.   The implications of these different behaviors and the impact they will have on observing are discussed more in Section \ref{sec:discus}.

For the IMRO test we read and saved the maximum $N =$ 256 reads per exposure. For this test the lens cover was on with additional aluminum covering. The temperature for each exposure was $-40^{\circ}$ C. We took one full frame exposure saving 256 raw reads with the following frame rates: 400, 200, 100, 50, 25, and 10 FPS, which correspond to exposure times of 0.64, 1.28, 2.56, 5.12, 10.24, and 25.6 seconds.

The right six panels in Figure \ref{fig:pixelcounts} shows the charge accumulation of the same grouping of pixels as the left panel except now in IMRO mode. Each panel follows the same pixels through IMRO exposures with different FPS. In the final panel (10 FPS) the normal pixels go non-linear after they reach a threshold level count level. We treat values above this threshold has effectively saturated and only use the non-saturated data.

\subsection{Dark Current}
\label{sec:dc}

\begin{figure}
\begin{center}
\begin{tabular}{c}
\includegraphics[width=.9\columnwidth]{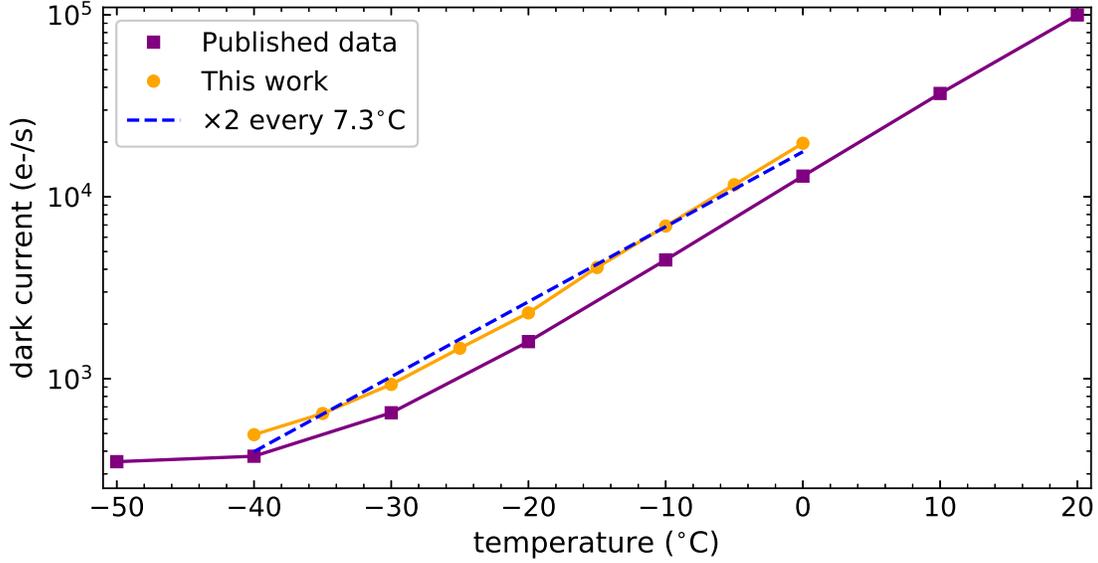}
\end{tabular}
\end{center}
\caption 
{ \label{fig:darkcurrent}
Measured IMRO dark current plus background radiation (orange circle) versus temperature of the detector, and previously published values from First Light (purple square). 
Each data point from this work is the average linear least squares fit of 255 non-destructive reads taken at a framerate of 400 FPS.
We excluded the first CDS read from the fit and masked out the small fraction of non-ideal pixels described in section \ref{sec:ndr}.
The blue dashed line is a fit to the average dark current plus background radiation.
} 
\end{figure}

We attempted to measure the C-RED 2 dark current using single full frame IMRO exposures at nine temperatures. For each exposure the lens cover was on with additional aluminum covering. We used a frame rate of 400 FPS and saved the maximum 256 reads. We sampled the detector at temperatures $T = -40, -35, -30, -25, -20, -15, -10, -5,$ and $0^{\circ}$ C. For each exposure we calculated the linear least squares fit of every pixel, excluding the small fraction of hot, non-ideal, or leaking pixels described in section \ref{sec:ndr}. The average charge accumulation rates for each temperature are in Figure \ref{fig:darkcurrent} along with the dark current values published by First Light in \citenum{2018SPIE10703E..1VF}.  Our accumulation rates are 1.3-1.5 times higher than those previously published. The current we measured most likely includes electrons from the 300 K black body background radiation in the lab because we did not use a 70 K cold stop. Using a solid angle of 70$^{\circ}$, which roughly corresponds to the angle created by the detector array and the camera opening, we calculated the black body radiation in Appendix \ref{sec:bb} to be 487 e$^-$/s, which is comparable our recorded dark current of 493 e$^-$/s at -40$^{\circ}$ C. While this is not the true dark current, we do find that the dark current plus background radiation we measure increases by a factor of 2 every 7.3$^{\circ}$ C, which is similar to rate published by First Light. We also note the current advertised data sheet for the C-RED 2 quotes a dark current of 600 e$^-$/s/pix.  

\subsection{Non-Destructive Read Noise} 
\label{sec:ndr}
Here we aim to quantify the noise of the integrated signal on the detector when using multiple non-destructive reads.  A series of NDRs can be used to create a charge accumulation rate for each pixel, which theoretically significantly reduces the readout noise compared to the individual reads, as first introduced in \citenum{1991SPIE.1541..127F} and expanded on in \citenum{1993SPIE.1946..395G} and \citenum{2007wfc..rept...12R}. Figure 6 in \citenum{2018SPIE10703E..1VF} shows the comparison between CDS noise and NDR noise of the C-RED 2. The noise of the slope, $\sigma_b$, is derived in Equation 1.33 of \citenum{2007wfc..rept...12R}:

\begin{equation}\label{eq:sigmab}
    \sigma_b = \sqrt{\frac{12}{N(N^2-1)}}\frac{\sigma_y}{dt} 
\end{equation}

\noindent Here $\sigma_y$ is the noise of the charge on the pixels at read $N$, and $dt$ is the constant time spacing between reads. They go on to derive the error of the integrated signal $\sigma_S$, which is the slope error multiplied by the integration time:
\begin{equation}\label{eq:NDRtint}
    \sigma_S = t_{\text{int}}\sigma_b 
\end{equation}
\noindent They use an integration time of $t_{\text{int}}=(N-1) dt$, which assumes the first frame read from the detector is subtracted from the subsequent images. In this work we do not subtract the initial frame, so the integration time is simply $t_{\text{int}}=N dt$ and the error of the integrated signal using Equations \ref{eq:sigmab} and \ref{eq:NDRtint} is:

\begin{equation}
    \sigma_S = \sigma_y \sqrt{\frac{12N}{(N^2-1)}}
\end{equation}

The total noise ($\sigma_{tot}$) we measure includes shot noise from the cumulative dark and background current ($i_{\text{DC+BB}}$, which we measured in Section \ref{sec:dc} to be 493 e$^-$/s). We also denote any other noise as {\it other}. The total noise measured can be expressed as: 

\begin{equation}
    \sigma_{tot} = \sqrt{\frac{12N}{(N^2-1)}\sigma_y^2 + \frac{6(N^2+1)(N-1)}{5N(N+1)}\tau i_{\text{DC+BB}} + {\text {other}}}
\end{equation}

Where $\tau$ is the time between reads (1/FPS). The total CDS noise ($\sigma_{\text{CDS}}$) can be expressed similarly: 

\begin{equation}
    \sigma_{\text{CDS}}=\sqrt{2\sigma_y^2+\tau i_{\text{DC+BB}}+\text{other}}
\end{equation}

\begin{figure}
\begin{center}
\begin{tabular}{c}
\includegraphics[width=\columnwidth]{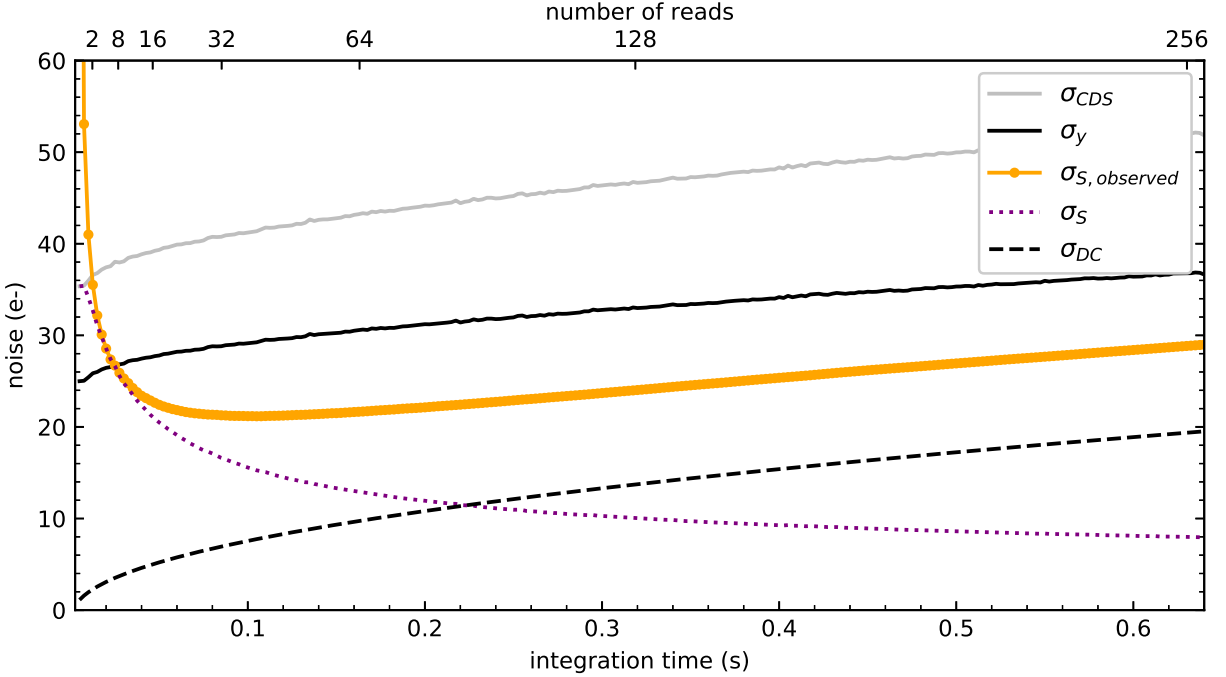}
\end{tabular}
\end{center}
\caption 
{ \label{fig:rnVread}
The theoretical noise of an image created with multiple non-destructive reads (dotted purple) as a function of image exposure time. The theoretical values are based on the average readout noise of the individual frames (solid black). The noise of the calculated slope images is shown in orange. Also plotted is the CDS noise for comparison (solid gray) and shot noise from the dark current and background photons (dashed black).
} 
\end{figure}

In this section we calculate the readout noise of 256 reads and compare it to the noise of the final integrated image. We recorded 256 reads from 10 dark, full frame exposures at 400 FPS and $T = -40^{\circ}$ C. We calculated the linear least squares fit from $N=2$ through $N=256$.  After calculating the charge accumulation rates for every pixel, we multiplied by the exposure time ($t_{\text{int}}=N/{\text{FPS}}$) to create the integrated signal image.  We then calculated the noise of each pixel between the 10 exposures for each of the 255 integrated arrays. The average noise of each array is plotted in Figure \ref{fig:rnVread}. Also shown in this figure is the theoretical noise of the final image, the average noise of each of the 256 reads, the noise due to the dark current and background radiation, and the CDS noise.

We find that the observed noise is within 2 e$^-$ of the theoretical noise between $N$ = 4 and $N$ = 18 reads.
The measured noise diverges from the read noise limit for longer exposures due to the added background noise.
Even with the added sources of noise, the measurements are comparable to Figure 6 in \citenum{2018SPIE10703E..1VF}. The lowest effective read noise we can achieve using up-the-ramp sampling is 21.2 e$^-$ at N = 42 reads at 400 FPS. This is half the noise from a single frame image with the same exposure time.

\section{Signal-to-Noise}
\label{sec:snr}
In this section we calculate the expected signal-to-noise ratio (SNR) of the C-RED 2 in using both CDS and IMRO images.  We do this to determine which exposure types are best suited for different target brightnesses, observing conditions, and exposure times. The SNR depends on the exposure time, shot noise, dark current plus background radiation electron rate ($N_{\text{DC+BB}}$), and read noise ($N_R$):
\begin{equation} \label{eq:snr}
{\text SNR} = \frac{N_*t_{\text{int}}}{\sqrt{N_*t_{\text{int}} + nN_{\text{DC+BB}} t_{\text{int}} + nN_R^2}}
\end{equation}
The number of pixels, $n$, is the aperture, which in this case is equal to the C-RED 2 point spread function of $5 \times 5$ pixels. We calculate the incident electron rate on the C-RED 2 from a target star, $N_*$, in Appendix \ref{sec:Nstar}.

\begin{figure}
\begin{center}
\begin{tabular}{c}
\includegraphics[width=.7\columnwidth]{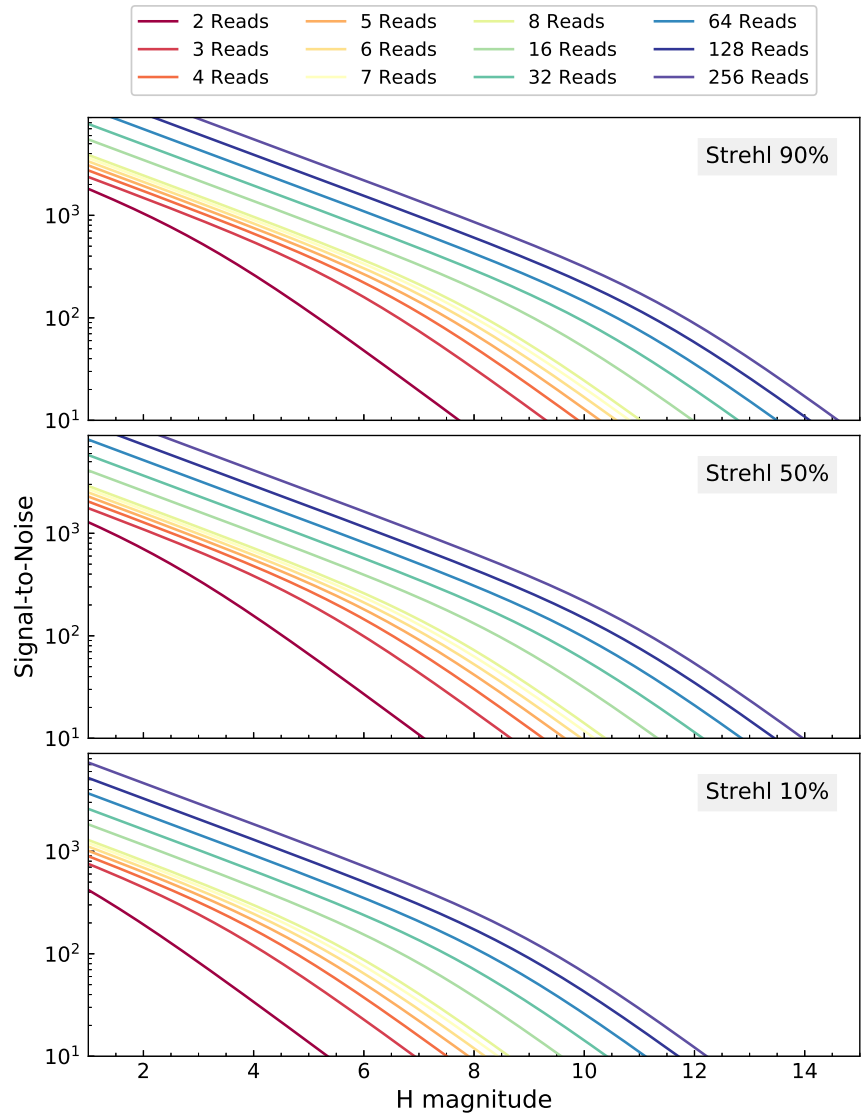}
\end{tabular}
\end{center}
\caption 
{ \label{fig:snrVh}
Signal-to-noise expected as a function of target star magnitude at Strehl ratios of 90, 50, and 10$\%$.  The SNR calculation includes shot noise, dark current plus background black body radiation, and read noise characterized in this work. Each line is the SNR calculated using multiple non-destructive reads collected in IMRO mode. 
} 
\end{figure}

For IMRO mode we will operate the C-RED 2 at 400 FPS, $T = -40^{\circ}$ C.  Therefore we calculate the SNR with a dark current and background radiation of 493 $e^-$/s and use the calculated NDR noise for the read noise. For CDS SNR we use $\sigma_{\text{CDS}}$ for the read noise term. The SNR values for IMRO mode at three different Strehl ratios are show in Figure \ref{fig:snrVh}.  The bend in the SNR in this figure show where the detector switches from shot nosie dominated (higher SNR) to read noise dominated (lower SNR). The transition occurs at higher SNR for lower read numbers. For example, at 2 reads the noise is completely read noise dominated except for the brightest of targets, whereas at 256 reads and 90$\%$ Strehl, the noise becomes shot noise dominated for stars brighter than H = 12 magnitude.

\begin{figure}
\begin{center}
\begin{tabular}{c}
\includegraphics[width=.8\columnwidth]{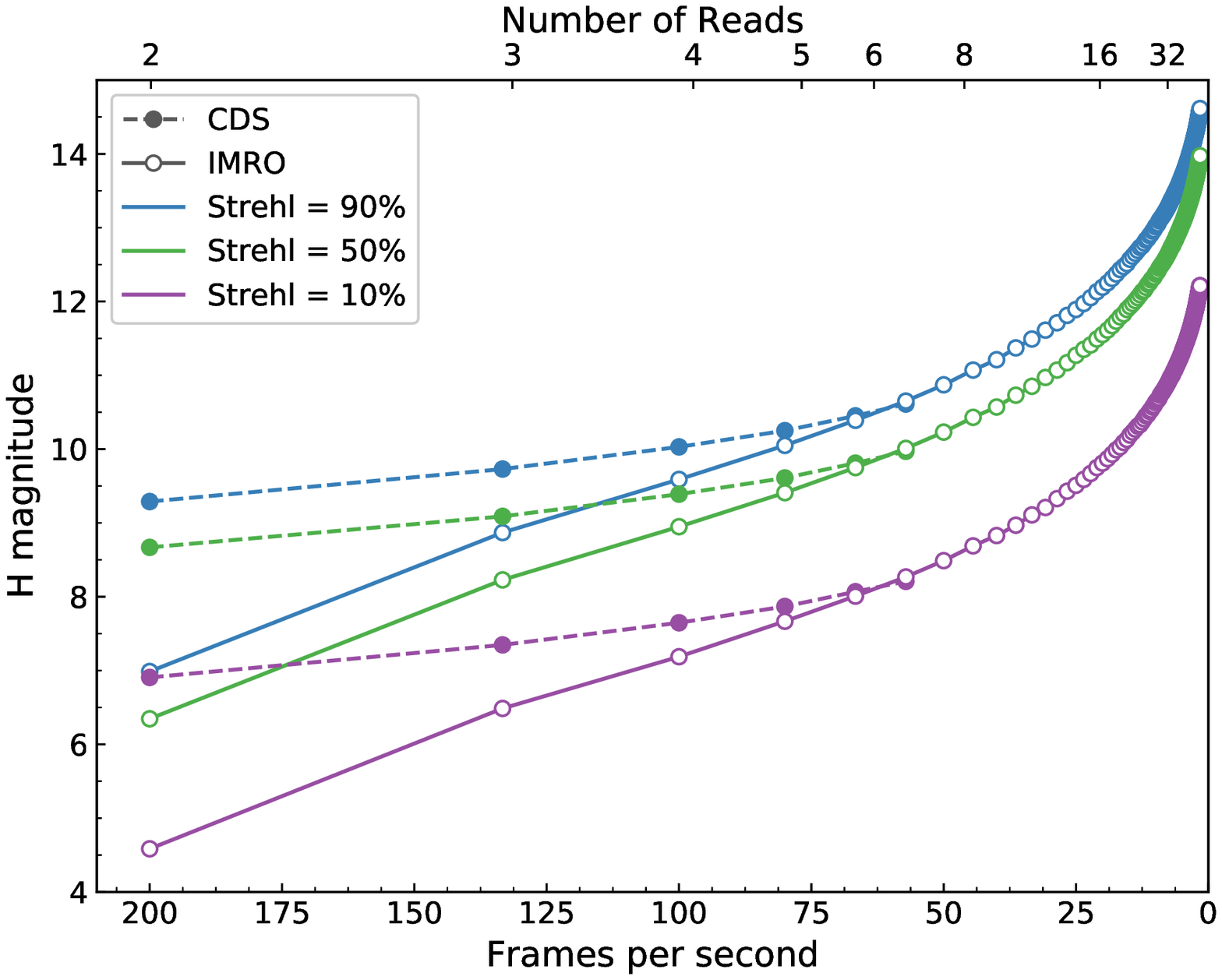}
\end{tabular}
\end{center}
\caption 
{ \label{fig:fpsVh}
Number of reads in IMRO mode (open circles), or framerate in CDS mode (filled circles) required to achieve a signal-to-noise of 10 for a given H magnitude at Strehls of 90, 50, and 10$\%$. This requires the camera to operate at 400 FPS in IMRO mode, and in both IMRO and CDS modes the detector is operating at $-40^{\circ}$ C.
} 
\end{figure}

Figure \ref{fig:fpsVh} shows the number of reads necessary in IMRO mode, and the frame rate necessary in CDS mode, to achieve a SNR of 10 for a given target star magnitude.  To clarify, when using IMRO mode we will always operate at a framerate of 400 FPS, so the exposure time, and thus the SNR, is determined by the number of reads.  To achieve the same exposure time in CDS mode we will vary the detector framerate and use the maximum integration time such that FPS = $400/N$ and $t_{\text{int}} = N/400$. These results will be key to operating the fast guiding loop because they illustrate where it becomes more efficient to use multiple reads instead of single CDS images. When guiding on brighter stars it is faster to use CDS mode.  More specifically when Strehl is 90, 50, and 10$\%$, we should use CDS mode for stars brighter than H magnitude of 10.3,  9.6, and  7.9, respectively.  For stars dimmer it is faster to use IMRO mode. We find that the limits to achieve a minimum SNR of 10 using the longest exposure time ($t_{\text{int}}=0.64$ seconds; $N=256$)  are $m_H$ = 14.6, 14.0, and 12.2 for S = 90, 50, and 10$\%$. 

\section{Conclusions}
\label{sec:discus} 

We find that the C-RED 2 performs optimally at $T=-40^{\circ}$C and 400 FPS with a dark current of 493 $e^-$/s. Using up-the-ramp sampling we are able to reduce the read out noise to a minimum of 21.2 e$^-$ using 42 samples, or 29.0 e$^-$ using the maximum 256 samples. The results presented here are consistent with those published by First Light in \citenum{2018SPIE10703E..1VF} and \citenum{2018SPIE10539E..14G} and those in the performance check we received with our C-RED 2.  An exception is the dark current, which is 1.3 times higher than the measurements published by First Light.  It is likely our measurements of the dark current includes light from the 300 K black body of the lab.

While most of the pixels have a linear response, it is worth noting that a small fraction of pixels exhibited non-ideal behavior in both CDS and IMRO mode.  The C-RED pixels can be classified into four types: normal charge accumulation ($99.984\%$ of pixels), non-ideal accumulation ($0.006\%$), leaking charge ($0.004\%$), and hot pixels ($0.006\%$). Examples of these are shown in Figure \ref{fig:pixelcounts}. Hot pixels are expected in detectors, however the non-ideal and leaking pixels that lose charge over time are unusual. The cause of this behavior is unknown, however the fraction of good to bad (non-ideal, leak, and hot) pixels is 6300 to 1, so the unusual behavior will have an inconsequential effect on the performance of the camera. Overall the detector is linear and within specifications.

Using the results of this characterization and optical design of PARVI, we calculated the expected signal-to-noise of the C-RED 2. With our SNR estimates we determined that when the tip/tilt loop is operating faster than 80 Hz, we can achieve a better signal using CDS images rather than IMRO images, because at that speed the IMRO read noise is much higher than the CDS noise. When the loop is operating at 80 Hz or slower, the NDR noise decreases significantly and we will use IMRO images instead of CDS.  We find that when using the optimal settings and this image configuration, the faintest target the C-RED 2 can guide on is a 14.6 H magnitude star. With the recent updates to P3K, the telescope is able to correct on targets as dim as this. Alternatively, the telescope can correct and lock on to a brighter target to acheive a higher Strehl, while the C-RED 2 guides on a fainter, nearby companion.

We have shown that the C-RED 2 operates within expectations and will perform well as the PARVI guide camera. We began commissioning PARVI in June 2019.

\appendix    

\section{Background Radiation}
\label{sec:bb}
We calculate the total power per unit area from a 300 K black body emitter between the operation wavelengths of the C-RED 2  ($\lambda = 0.9 - 1.7$ $\rm{\mu}$m; $\nu = 1.8 - 3.3 \times 10^{14}$ Hz)  with Planck's Law:
\begin{equation}
P = \Omega \int \frac{2h\nu^3}{c^2} \text{exp}\left[\frac{h\nu}{kT}-1\right]
\end{equation}
Here we used a solid angle of $\Omega = 1.22$ radians, which is the solid angle between the detector and the camera opening. Using the energy of a single photon of wavelength $1.65$ $\rm{\mu}$m ($E_{1.65\rm{\mu}\text{m}}$), and the area of a single pixel ($A_{\text{pix}}=225 \rm{\mu}\text{m}^2$), we calculate the rate of photons on each pixel ($N_e$):
\begin{equation}
N_e = P \times A_{\text{pix}} \times E_{1.65\rm{\mu}\text{m}}
\end{equation}
Using the Quantum Efficiency ($QE=$70$\%$) we find the count rate on the detector from background radiation, $N_{BB}$:
\begin{equation}
N_{BB} = N_e \times QE
\end{equation}
This yields a background electron rate of  487.4 $e^-$/s.

\section{Incident Stellar Electron Rate}
\label{sec:Nstar}
For a given apparent H magnitude, $m_h$, the flux received by the telescope, $F_h$, is:
\begin{equation}
F_h = F_0  10^{-m_h/2.5}
\end{equation}
where $F_0 =$ ($1.86\times 10^{-6}$ erg/s/Hz) is the H band flux of Vega at $\lambda=1.65$  $\rm{\mu m}$ \cite{2003AJ....126.1090C}. The energy that enters the adaptive optics given the area of the primary ($A_{prime}$) and secondary ($A_{second}$) mirrors (81.71 m$^2$ and 3.45 m$^2$, respectively) is: 
\begin{equation}
E_h = \left(A_{prime}-A_{second}\right) F_h 
\end{equation}
The total electron signal on the detector from the star depends on the throughput to the C-RED 2 ($\eta = 1\%$), the Quantum Efficiency of the detector ($QE = 70\%$), Strehl ratio of the optics, $S$, and energy of a photon of wavelength $\lambda =1.65$ $\rm{\mu m}$ ($E_{1.65\mu{\text{m}}}$) as such:
\begin{equation}
N_* =  S \left(\frac{\eta  QE }{E_{1.65\mu{\text{m}}}} \right) E_h 
\end{equation}

\subsection*{Disclosures}
The authors have no relevant financial interests in this manuscript nor any other potential conflicts of interest.

\acknowledgments 
RKG is supported by a Kalbfleisch Graduate Fellowship at the American Museum of Natural History.  

\bibliography{main}   
\bibliographystyle{spiejour}   


\vspace{2ex}\noindent\textbf{Rose K. Gibson} is a doctoral candidate in Astronomy at Columbia University and funded by the American Museum of Natural History in New York. She received her BA in Astrophysics from Wellesley College in 2016, and has received both her MA and MPhil in Astronomy from Columbia University in 2018 and 2019, respectively. She is currently working on the Palomar Radial Velocity Instrument data acquisition program, data reduction pipeline, and survey target selection.

\begin{sidewaystable}
\begin{center} \footnotesize
\begin{tabular}{|p{1.9cm}|c|c|p{1.8cm}|p{1.7cm}|p{5cm}|p{1.9cm}|p{4.1cm}|} \hline
Instrument	& Telescope  & Aperture (m) & First Light / Availability & Predicted NIR-RV instrument precision & Spectral Resolution & Wavelength Reference & Reference \\ \hline
CARMENES & Calar Alto & 3.5 & 01/2016 Facility & 1 ms$^{-1}$ & R=94,600, 80,400 (520-960 nm, 960-1710 nm) stabilized fiber-fed spectrographs & Etalon & Quirrenbach et al. 2016\cite{2016SPIE.9908E..12Q}, 2018\cite{2018SPIE10702E..0WQ} \\ \hline
GIARPS (HARPS-N + GIANO-B) & TNG & 3.6 & 03/2017 Facility & 3 ms$^{-1}$ & R=115,000, 50,000 (383-2450 nm)  stabilized spectrograph & Th-Ar lamp & Claudi et al. 2016\cite{2016SPIE.9908E..1AC} \\ \hline
HPF & HET & 10.0 & 11/2017 Facility & 1 ms$^{-1}$ & R=50,000 (818-1700 nm) stabilized fiber-fed spectrograph & Laser Frequency Comb & Mahadevan et al. 2014\cite{2014SPIE.9147E..1GM} \\ \hline
{\bf iLocator} & LBT & 2$\times$8.4 & 10/2019 PI-only & 0.4 ms$^{-1}$ & R=150,000-240,000 (970-1300 nm) stabilized SM fiber-fed spectrograph with full AO correction & Etalon & Crepp et al. 2016\cite{2016SPIE.9908E..19C} \\ \hline
{\bf IRD} & Subaru & 8.2 & 10/2017 Facility & $\sim$1 ms$^{-1}$ & R=70,000 or 100,000 (970-1750 nm) ceramic component stabilized, partial AO correction, MMF or SMF spectrograph & Laser Frequency Comb & Kotani et al. 2018\cite{2018SPIE10702E..11K} \\ \hline
iSHELL & IRTF & 3.0 & 10/2016 Facility & $<$10 ms$^{-1}$ & R$\sim$75,000, Cassegrain mounted spectrograph (1100-5300 nm) & Gas cell and lamp & Rayner et al. 2016\cite{2016SPIE.9908E..84R} \\ \hline
NIRPS & La Silla & 3.6 & 08/2019 & 1 ms$^{-1}$ & R$\sim$100,000 (950-1800 nm) stabilized fiber-fed spectrograph with partial (0.4'') AO correction & Lamp and Fabry-Perot & Wildi et al. 2017\cite{2017SPIE10400E..18W}	\\ \hline
{\bf PARVI} & Palomar & 5.1 & 6/2019 PI & 30 cms$^{-1}$ & R$\sim$100,000 (1250-1800 nm) stabilized SM fiber-fed spectrograph with full extreme AO correction & Laser Frequency Comb & Internal \\ \hline
SPIROU & CFHT & 3.6 & 03/2018 Facility & 1 ms$^{-1}$ & R$\sim$70,000 (980-2440 nm) stabilized fiber-fed spectrograph & Gas cell and Fabry-Perot & Donati et al. 2018\cite{2017haex.bookE.107D} \\ \hline
\end{tabular}
\end{center}
\caption{\label{tab:apra}
Planned radial velocity instruments operating beyond 1 micron wavelength. For a comprehensive list see \citenum{2017RNAAS...1a..51W}. Instruments in bold are those using AO correction and SMFs. 
}
\end{sidewaystable}

\listoffigures

\end{document}